\documentclass[reprint,amsmath,amssymb,aps,superscriptaddress,nofootinbib]{revtex4-1}

\usepackage[utf8]{inputenc}
\usepackage[T1]{fontenc}

\usepackage{tgtermes}
\usepackage{newtxmath}
\usepackage{microtype}

\usepackage{mathtools}
\usepackage{bm}

\usepackage{graphicx}
\usepackage{rotating}

\usepackage{dcolumn}
\usepackage{booktabs}
\usepackage{siunitx}

\usepackage[svgnames]{xcolor}
\usepackage{soul}
\usepackage{appendix}
\usepackage{orcidlink}
\usepackage{comment}

\definecolor{accent}{RGB}{135,105,175}

\usepackage{hyperref}

\hypersetup{
    colorlinks = true,
    linkcolor  = accent,
    citecolor  = accent,
    urlcolor   = accent
}

\begin{document}

\title{Dymnikova black string}

\author{L. G. Barbosa \orcidlink{0009-0007-3468-3718}}
\email{leonardo.barbosa@posgrad.ufsc.br}
\affiliation{Departamento de Física, CFM - Universidade Federal de \\ Santa Catarina; C.P. 476, CEP 88.040-900, Florianópolis, SC, Brazil}

\author{L. C. N. Santos \orcidlink{0000-0002-6129-1820}}
\email{luis.santos@ufsc.br}
\affiliation{Departamento de Física, CFM - Universidade Federal de \\ Santa Catarina; C.P. 476, CEP 88.040-900, Florianópolis, SC, Brazil}

\author{C. C. Barros Jr. \orcidlink{0000-0003-2662-1844}}
\email{barros.celso@ufsc.br }
\affiliation{Departamento de Física, CFM - Universidade Federal de \\ Santa Catarina; C.P. 476, CEP 88.040-900, Florianópolis, SC, Brazil}

\begin{abstract}
We obtain a new static black string solution of Einstein's equations with a negative cosmological constant sourced by a Dymnikova-type anisotropic matter distribution. At large radial distances, the geometry approaches the Lemos black string, while the behavior near the central region is characterized by an effective curvature parameter. We examine the weak and strong energy conditions, the horizon structure, and the curvature invariants. The Ricci and Kretschmann scalars remain finite at the origin. We also investigate the Hawking temperature and heat capacity, with particular attention to the dependence of the thermodynamic behavior on the Dymnikova scale.
\end{abstract}

\maketitle

\section{Introduction}

Black-string geometries provide a useful framework for investigating gravitational configurations with cylindrical symmetry, particularly in the presence of a negative cosmological constant. In four-dimensional general relativity, Lemos obtained a static cylindrically symmetric black-hole solution for $\Lambda<0$~\cite{Lemos:1994xp}. This construction was subsequently extended to include rotation and electric charge~\cite{Lemos:1995cm}. In the static case, the corresponding line element can be written as
\begin{equation}
	ds^{2}=-\left(\frac{r^{2}}{\ell^{2}}-\frac{2M}{r}\right)dt^{2}
	+\frac{dr^{2}}
	{\left(\frac{r^{2}}{\ell^{2}}-\frac{2M}{r}\right)}
	+r^{2}d\phi^{2}
	+\frac{r^{2}}{\ell^{2}}dz^{2},
\end{equation}
where $\ell$ denotes the AdS radius, $M$ is the mass parameter, and $r$ is the cylindrical radial coordinate.

Since these early constructions, black strings have been considered in a variety of gravitational and matter configurations. Static black strings surrounded by anisotropic quintessence were studied in Ref.~\cite{Ali:2019mxs}, while configurations involving string clouds and anisotropic quintessence were considered in Ref.~\cite{Deglmann:2025mcl}. Charged and rotating black strings in the presence of Kiselev-type matter have also been investigated~\cite{Barbosa:2025scy}, together with black strings immersed in perfect-fluid dark matter~\cite{Barbosa:2026guu}. Related configurations have been examined in $f(R,T)$ gravity~\cite{Santos:2026bjq}, in Lifshitz spacetime~\cite{Lessa:2024gbd}, and in charged black-string geometries containing string-cloud and quintessence contributions~\cite{EslamPanah:2026cjx}. These developments illustrate the range of matter sources and gravitational frameworks in which cylindrical black-hole geometries can be studied.

Black-string backgrounds have also been used to investigate field propagation and perturbative phenomena. The quantization of massive scalar fields and the vacuum polarization of massive spinor fields in static black-string spacetimes were analyzed in Refs.~\cite{FernandezPiedra:2007ujk,Piedra:2007yi}, while vacuum-polarization effects for scalar fields in cylindrical black-hole backgrounds were considered in Ref.~\cite{DeBenedictis:1998be}. Quasinormal modes have been studied for AdS black strings~\cite{Mirekhtiary:2022juf} and for black strings in de Rham--Gabadadze--Tolley massive gravity~\cite{Ponglertsakul:2018smo}, with greybody factors in the latter setting investigated in Ref.~\cite{Boonserm:2019mon}. Rotating black strings have furthermore appeared in holographic constructions~\cite{Brihaye:2010ce}. Their thermodynamic properties have also been considered in charged and rotating configurations, including extended phase-space analyses~\cite{Bakhtiarizadeh:2025uks,Zhen:2025mqz}.

Another aspect of interest concerns the behavior of cylindrical black-hole geometries in the central region. Regular black-string configurations have, for example, been investigated through black-bounce constructions~\cite{Lima:2023jtl}. More generally, the modification of central curvature divergences has motivated the study of regular black-hole geometries in several gravitational settings, including higher-curvature and Lovelock theories~\cite{Konoplya:2024kih,Estrada:2024uuu}. Among the earlier constructions in this context is the nonsingular black-hole geometry proposed by Dymnikova~\cite{Dymnikova:1992ux}.

The Dymnikova solution was originally formulated within spherical symmetry and is associated with an anisotropic matter distribution whose density remains finite at the origin~\cite{Dymnikova:1992ux}. Dymnikova-type geometries have subsequently been examined in higher-dimensional settings~\cite{Paul:2023pqn,Macedo:2024dqb}, in connection with higher-curvature corrections~\cite{Konoplya:2024kih}, and through analyses of tidal forces~\cite{Macedo:2026ehd}. Related regular configurations combining Dymnikova-type matter with other sources have also been investigated from thermodynamic and perturbative perspectives~\cite{Santos:2026xtf}. These studies motivate examining how a similar matter profile influences a spacetime with cylindrical rather than spherical symmetry.

To distinguish the spherical radial coordinate from the cylindrical coordinate employed below, we denote the former by $\bar r$. In the parametrization adopted here, the corresponding spherical geometry can be written as
\begin{equation}
	\begin{aligned}
		ds^{2}={}&-\left(1-\frac{2M D(\bar r)}{\bar r}\right)dt^{2}
		+\frac{d\bar r^{\,2}}
		{\left(1-\frac{2M D(\bar r)}{\bar r}\right)}
		+\bar r^{\,2}d\Omega^{2},
	\end{aligned}
\end{equation}
where $d\Omega^{2}=d\theta^{2}+\sin^{2}\theta\,d\phi^{2}$. The radial dependence of the mass contribution is encoded in the profile function
\begin{equation}\label{Dymnikova_function}
	D\left(\bar r\right)
	=1-e^{-\bar r^{3}/\bar r_{0}^{3}},
\end{equation}
where $\bar r_{0}$ determines the characteristic scale associated with the matter distribution. The profile modifies the mass contribution in the central region, while $D(\bar r)\rightarrow1$ for $\bar r\gg\bar r_0$.

Motivated by this structure, in the present work we consider a static cylindrically symmetric configuration in Einstein gravity with a negative cosmological constant and a Dymnikova-type anisotropic matter distribution. The radial profile is incorporated into the mass contribution of the black-string geometry, leading to the line element
\begin{equation}
	ds^{2}=-\left(\frac{r^{2}}{\ell^{2}}
	-\frac{2M D(r)}{r}\right)dt^{2}
	+\frac{dr^{2}}
	{\left(\frac{r^{2}}{\ell^{2}}
		-\frac{2M D(r)}{r}\right)}
	+r^{2}d\varphi^{2}
	+\frac{r^{2}}{\ell^{2}}dz^{2}.
\end{equation}
In the cylindrical configuration, $r_0$ controls the radial scale over which the Dymnikova-type contribution modifies the metric. For $r\gg r_0$, the exponential correction is suppressed and the geometry approaches the static Lemos black-string form. In the central region, the same profile alters the usual $1/r$ mass contribution and changes the leading behavior of the metric.

The aim of the present analysis is to examine the consequences of this matter distribution for the geometrical and thermodynamic properties of the black string. We derive the corresponding metric from Einstein's field equations, determine the transverse pressure from the conservation equation, and study the limiting behavior of the resulting geometry. We then analyze the weak and strong energy conditions, investigate the event-horizon structure, and evaluate the Ricci and Kretschmann scalars in order to characterize the central region. Finally, we examine the Hawking temperature and heat capacity, with particular attention to the role of the characteristic scale $r_0$.

The paper is organized as follows. In Sec.~\ref{Field_equation}, we derive the black-string solution from Einstein's field equations and discuss its limiting behavior. In Sec.~\ref{sec:energy}, we analyze the corresponding energy conditions. The event-horizon structure and curvature invariants are examined in Sec.~\ref{sec:horizons-singularities}. In Sec.~\ref{sec:thermodynamics}, we investigate the Hawking temperature and heat capacity of the solution. Finally, Sec.~\ref{Conclusions} summarizes the main results and discusses possible extensions.

\section{Field equations}\label{Field_equation}

In this section, we construct a static black string solution sourced by a Dymnikova-type anisotropic matter distribution in the presence of a negative cosmological constant. We begin by specifying the geometry and the matter content, and then use the conservation of the energy-momentum tensor to determine the transverse pressure. Once the matter sector is completely characterized, Einstein's field equations reduce to a set of radial equations for the metric function $f(r)$. Their solution and its limiting behavior are discussed subsequently.

\subsection{Einstein equation}

We consider Einstein's field equations with a negative cosmological constant, adopting $G=c=1$ and the signature $(-,+,+,+)$:
\begin{equation}
	\mathcal{G}_{\mu \nu}\equiv R_{\mu\nu}-\frac{1}{2}Rg_{\mu\nu}+\Lambda g_{\mu\nu}
	=8\pi T_{\mu\nu},
\end{equation}
where $R_{\mu\nu}$ is the Ricci tensor, $R$ is the curvature scalar, $g_{\mu\nu}$ is the metric tensor, and $T_{\mu\nu}$ is the energy-momentum tensor. The cosmological constant is parametrized as $\Lambda=-3/\ell^2$, where $\ell$ denotes the AdS radius.

For a static configuration with cylindrical symmetry, we adopt the line element
\begin{equation}\label{eq:metric}
	ds^{2}=-f\left(r\right)dt^{2}+\frac{dr^{2}}{f\left(r\right)}+r^{2}d\phi^{2}+\frac{r^{2}}{\ell^{2}}dz^{2},
\end{equation}
with $-\infty<t<+\infty$, $0\leq r<+\infty$, $0\leq\phi<2\pi$, and $-\infty<z<+\infty$. Within this ansatz, the radial dependence of the geometry is entirely encoded in the metric function $f(r)$.

The matter source is modeled as an anisotropic fluid described by the mixed energy-momentum tensor
\begin{equation}
	{T^{\mu}}_\nu=\text{diag}\left(-\rho,p_{r},p_{\perp},p_{\perp}\right),
\end{equation}
where $\rho$ is the energy density, $p_r$ is the radial pressure, and $p_{\perp}$ denotes the pressure along the two transverse directions. The matter variables are constrained by the conservation equation
\begin{equation}
	\nabla_{\mu}{T^{\mu}}_\nu=0.
\end{equation}
For the metric above, its nontrivial radial component reduces to
\begin{equation}
	p_{r}^{'}+\frac{f^{'}}{2f}\left(\rho+p_{r}\right)+\frac{2}{r}\left(p_{r}-p_{\perp}\right)=0.
\end{equation}
We now impose $p_r=-\rho$, as in the Dymnikova construction. Under this condition, the term containing $f'/f$ vanishes identically, and the conservation equation directly determines the transverse pressure as
\begin{equation}
	p_{\perp}=-\rho-\frac{r}{2}\rho^{'}. 
\end{equation}

We consider a Dymnikova-type energy-density profile~\cite{Dymnikova:1992ux}, given by
\begin{equation}
	\rho\left(r\right)=\frac{3M}{4\pi r_{0}^{3}}e^{-\frac{r^{3}}{r_{0}^{3}}},
\end{equation}
where $r_0$ sets the characteristic length scale of the matter distribution. This density remains finite in the central region and is exponentially suppressed for $r\gg r_0$.

Substituting this density into the conservation relation yields
\begin{equation}
	p_{\perp}=\frac{3M}{4\pi r_{0}^{3}}e^{-\frac{r^{3}}{r_{0}^{3}}}\left(\frac{3r^{3}}{2r_{0}^{3}}-1\right).
\end{equation}
The matter sector is therefore completely specified by $\rho$, $p_r=-\rho$, and $p_{\perp}$, allowing us to proceed to the gravitational field equations.

For the metric ansatz introduced above, the nonvanishing components of the Einstein tensor, including the cosmological contribution through $\mathcal{G}_{\mu\nu}$, are
\begin{align}
	\mathcal{G}_{tt} &= -f(r)\left(\frac{f'(r)}{r} + \frac{f(r)}{r^{2}} - \frac{3}{\ell^{2}}\right), \\
	\mathcal{G}_{rr} &= \frac{1}{f(r)}\left(\frac{f'(r)}{r} + \frac{f(r)}{r^{2}} - \frac{3}{\ell^{2}}\right), \\
	\mathcal{G}_{\phi\phi} &= r^{2}\left(\frac{1}{2}f''(r) + \frac{f'(r)}{r} - \frac{3}{\ell^{2}}\right), \\
	\mathcal{G}_{zz} &= \frac{1}{\ell^{2}}\,\mathcal{G}_{\phi\phi}.
\end{align}
On the matter side, the corresponding covariant components are
\begin{align}
	8\pi T_{tt} &= \frac{6Mf\left(r\right)}{r_{0}^{3}}e^{-\frac{r^{3}}{r_{0}^{3}}}, \\
	8\pi T_{rr} &= -\frac{6M}{r_{0}^{3}f\left(r\right)}e^{-\frac{r^{3}}{r_{0}^{3}}}, \\
	8\pi T_{\phi\phi} &= \frac{M}{r_{0}^{6}}e^{-\frac{r^{3}}{r_{0}^{3}}}\left(9r^{5}-6r^{2}r_{0}^{3}\right), \\
	8\pi T_{zz} &= \frac{M}{\ell^{2}r_{0}^{6}}e^{-\frac{r^{3}}{r_{0}^{3}}}\left(9r^{5}-6r^{2}r_{0}^{3}\right).
\end{align}

Equating the corresponding components of Einstein's equations gives
\begin{align}
	\frac{f^{'}}{r}+\frac{f}{r^{2}}-\frac{3}{\ell^{2}} &= -\frac{6M}{r_{0}^{3}}e^{-\frac{r^{3}}{r_{0}^{3}}}\\
	\frac{r^{2}}{2}f^{''}+rf^{'}-\frac{3}{\ell^{2}}r^{2} &= \frac{M}{r_{0}^{6}}e^{-\frac{r^{3}}{r_{0}^{3}}}\left(9r^{5}-6r^{2}r_{0}^{3}\right).
\end{align}
The first equation is first order in $f(r)$ and is sufficient to determine the metric function up to an integration constant. Once this equation is solved with the appropriate regularity condition, direct substitution shows that the resulting metric function also satisfies the second equation.

\subsection{Solution}

With the matter sector fully specified, the first radial field equation can be solved directly for the metric function. The resulting solution takes the form
\begin{equation}
	f\left(r\right)=\frac{r^{2}}{\ell^{2}}-\frac{2M}{r}\left(1-e^{-r^{3}/r^{3}_{0}}\right).
\end{equation}

This expression displays directly the two contributions to the geometry. The term $r^2/\ell^2$ is associated with the negative cosmological constant, whereas the second term contains the effect of the Dymnikova matter distribution. The exponential factor is relevant in the region controlled by $r_0$ and suppresses the usual $1/r$ behavior as the origin is approached.
The large-radius behavior follows from the exponential suppression of the matter contribution. For $r\gg r_0$, one obtains
\begin{equation}
	f(r)=\frac{r^2}{\ell^2}-\frac{2M}{r}
	+O\!\left(\frac{e^{-r^3/r_0^3}}{r}\right).
\end{equation}
Thus, at large distances, the metric approaches the asymptotic Lemos black string form.

Near the origin, the matter contribution modifies the leading behavior of the metric function. Introducing the effective curvature parameter
\begin{equation}\label{eq:leff}
	\frac{1}{\ell_{\mathrm{eff}}^{2}}
	\equiv
	\frac{1}{\ell^{2}}-\frac{2M}{r_0^{3}},
\end{equation}
the near-origin expansion can be written compactly as
\begin{equation}
	f(r)=\frac{r^{2}}{\ell_{\mathrm{eff}}^{2}}+O(r^5).
\end{equation}
The sign of $\ell_{\mathrm{eff}}^{-2}$ therefore determines the leading curvature character of the core.

Equation~\eqref{eq:leff} shows that the Dymnikova matter distribution modifies the coefficient of the leading quadratic term in the metric near the origin. In contrast to the vacuum Lemos solution, where this coefficient is determined solely by the AdS scale $\ell$, here it depends on both the cosmological and matter contributions. The effective curvature parameter $\ell_{\mathrm{eff}}^{-2}$ may therefore be positive, zero, or negative, depending on the values of $M$, $r_0$, and $\ell$.

Consequently, depending on the relative magnitude of $2M/r_0^3$ and $1/\ell^2$, the matter contribution can preserve, cancel, or reverse the sign of the quadratic term. The resulting local geometries are summarized in Table~\ref{tab:regimes}. This classification describes only the behavior of the metric near the origin and does not determine the horizon structure of the spacetime.

\begin{table}[htbp]
	\caption{Near-origin regimes of the solution according to the effective curvature parameter defined in Eq.~\eqref{eq:leff}.}
	\label{tab:regimes}
	\centering
	\begin{tabular}{lll}
		\toprule
		Condition & $\ell_{\mathrm{eff}}^{-2}$ & Core geometry\\
		\midrule
		$2M\ell^2<r_0^3$ & $>0$ & AdS-like\\
		$2M\ell^2=r_0^3$ & $=0$ & Vanishing curvature\\
		$2M\ell^2>r_0^3$ & $<0$ & de Sitter-like\\
		\bottomrule
	\end{tabular}
\end{table}

\begin{figure}[htbp]
	\centering
	\includegraphics[width=0.92\columnwidth]{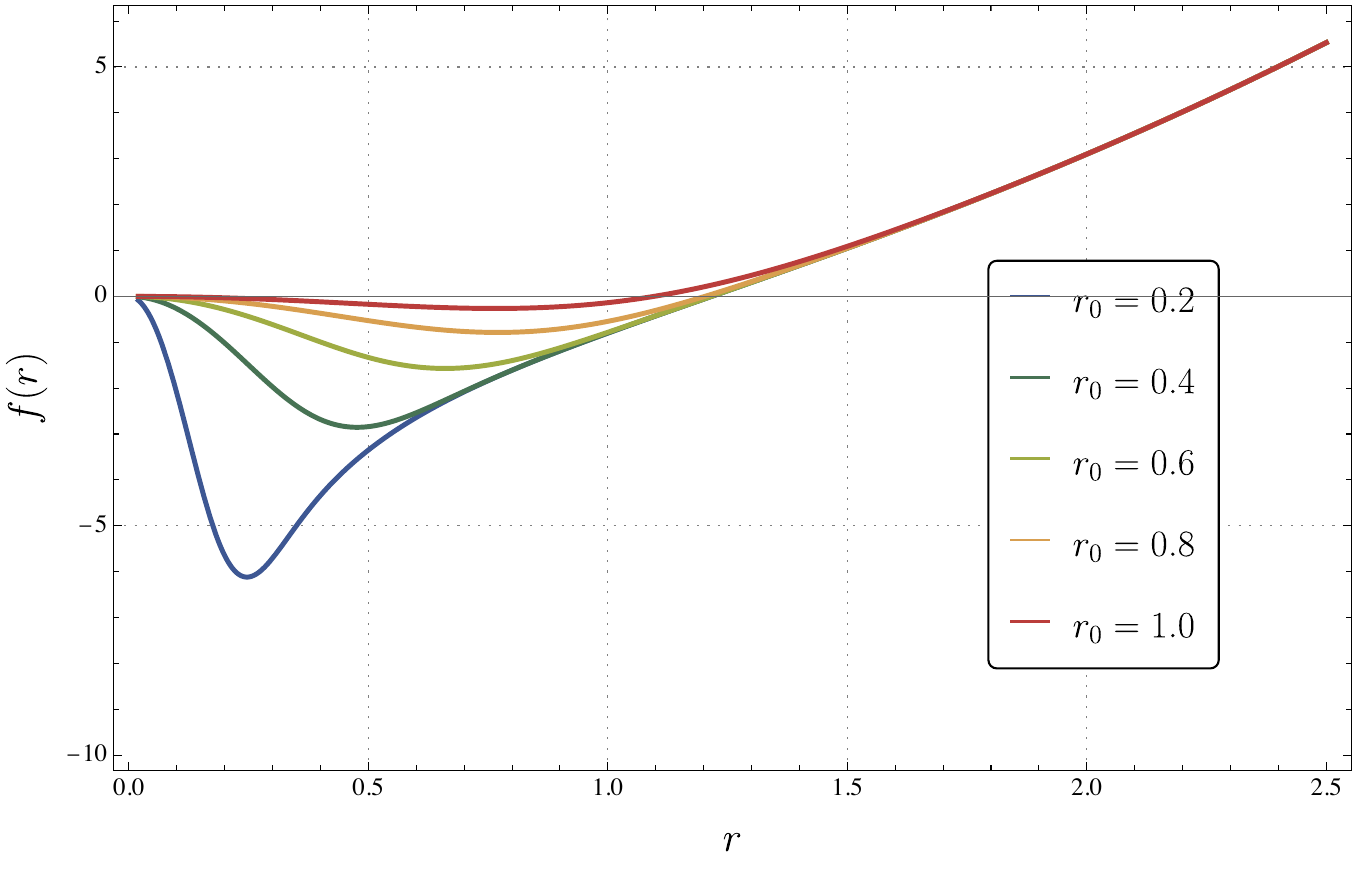}
	\caption{Radial behavior of the metric function $f(r)$ for $M=0.9$ and $\ell=1$, considering the representative values $r_0=0.2$, $0.4$, $0.6$, $0.8$, and $1.0$ of the Dymnikova scale.}
	\label{fig:f_function}
\end{figure}

The radial behavior of the metric function $f(r)$ illustrates how the Dymnikova scale affects the geometry between the central and asymptotic regions. Figure~\ref{fig:f_function} shows $f(r)$ for representative values of $r_0$, with $M$ and $\ell$ held fixed.

\section{Energy conditions}\label{sec:energy}

Having determined the matter distribution supporting the geometry, we now analyze the corresponding energy conditions. Throughout this section, we assume $M>0$, $r_0>0$, $\ell>0$, and $r>0$. Although the cosmological constant is kept on the geometric side of Einstein's equations, in the energy-condition analysis we include it as an effective contribution to the total energy density and pressures. For $\Lambda=-3/\ell^2$, the effective cosmological contribution is
\begin{equation}
	\rho_{\Lambda}=-\frac{3}{8\pi\ell^2},
	\qquad
	p_{r,\Lambda}=p_{\perp,\Lambda}
	=\frac{3}{8\pi\ell^2}.
\end{equation}

Combining these terms with the Dymnikova matter distribution, the effective energy density and principal pressures are
\begin{align}
	\rho_{\mathrm{eff}}
	&=
	\frac{3M}{4\pi r_0^3}
	e^{-r^3/r_0^3}
	-\frac{3}{8\pi\ell^2},
	\\
	p_{r,\mathrm{eff}}
	&=
	-\frac{3M}{4\pi r_0^3}
	e^{-r^3/r_0^3}
	+\frac{3}{8\pi\ell^2},
	\\
	p_{\perp,\mathrm{eff}}
	&=
	\frac{3M}{4\pi r_0^3}
	e^{-r^3/r_0^3}
	\left(
	\frac{3r^3}{2r_0^3}-1
	\right)
	+\frac{3}{8\pi\ell^2}.
\end{align}

The effective radial pressure retains the relation
$p_{r,\mathrm{eff}}=-\rho_{\mathrm{eff}}$, so that
$\rho_{\mathrm{eff}}+p_{r,\mathrm{eff}}=0$. In the transverse direction, the cosmological contribution cancels and one obtains
\begin{equation}\label{eq:rho-pperp-sum}
	\rho_{\mathrm{eff}}+p_{\perp,\mathrm{eff}}
	=
	\frac{9Mr^3}{8\pi r_0^6}
	e^{-r^3/r_0^3}
	\geq0.
\end{equation}

Thus, the combinations involving the principal pressures are automatically nonnegative for the parameter range considered here. The remaining restrictions associated with the weak and strong energy conditions are analyzed below.

\subsection{Weak energy condition}

The weak energy condition (WEC) requires the effective energy density to be nonnegative, together with the corresponding inequalities involving the principal pressures,
\begin{equation}\label{eq:wec-definition}
	\rho_{\mathrm{eff}}\geq0,
	\qquad
	\rho_{\mathrm{eff}}+p_{r,\mathrm{eff}}\geq0,
	\qquad
	\rho_{\mathrm{eff}}+p_{\perp,\mathrm{eff}}\geq0.
\end{equation}

The last two inequalities have already been satisfied. Therefore, the only nontrivial restriction comes from the sign of the effective energy density. Substituting $\rho_{\mathrm{eff}}$ gives
\begin{equation}\label{eq:wec}
	\frac{2M}{r_0^3}
	e^{-r^3/r_0^3}
	-\frac{1}{\ell^2}
	\geq0.
\end{equation}

The Dymnikova contribution in Eq.~\eqref{eq:wec} is positive and decreases monotonically with the radial coordinate, whereas the cosmological contribution remains constant. A positive radial region satisfying the WEC therefore exists only when
\begin{equation}\label{eq:wec-parameter}
	2M\ell^2>r_0^3.
\end{equation}
In this case, Eq.~\eqref{eq:wec} can be solved explicitly, yielding
\begin{equation}\label{eq:wec-range}
	0<r\leq
	r_0
	\left[
	\ln\left(
	\frac{2M\ell^2}{r_0^3}
	\right)
	\right]^{1/3}.
\end{equation}

Thus, the WEC is satisfied only within a finite radial region and is violated beyond the upper bound in Eq.~\eqref{eq:wec-range}. If $2M\ell^2\leq r_0^3$, the WEC is violated for every $r>0$. At large distances, this behavior follows from the exponential suppression of the Dymnikova contribution, leaving the negative effective energy density associated with the cosmological constant.

\begin{figure}[htbp]
	\centering
	\includegraphics[width=\columnwidth]{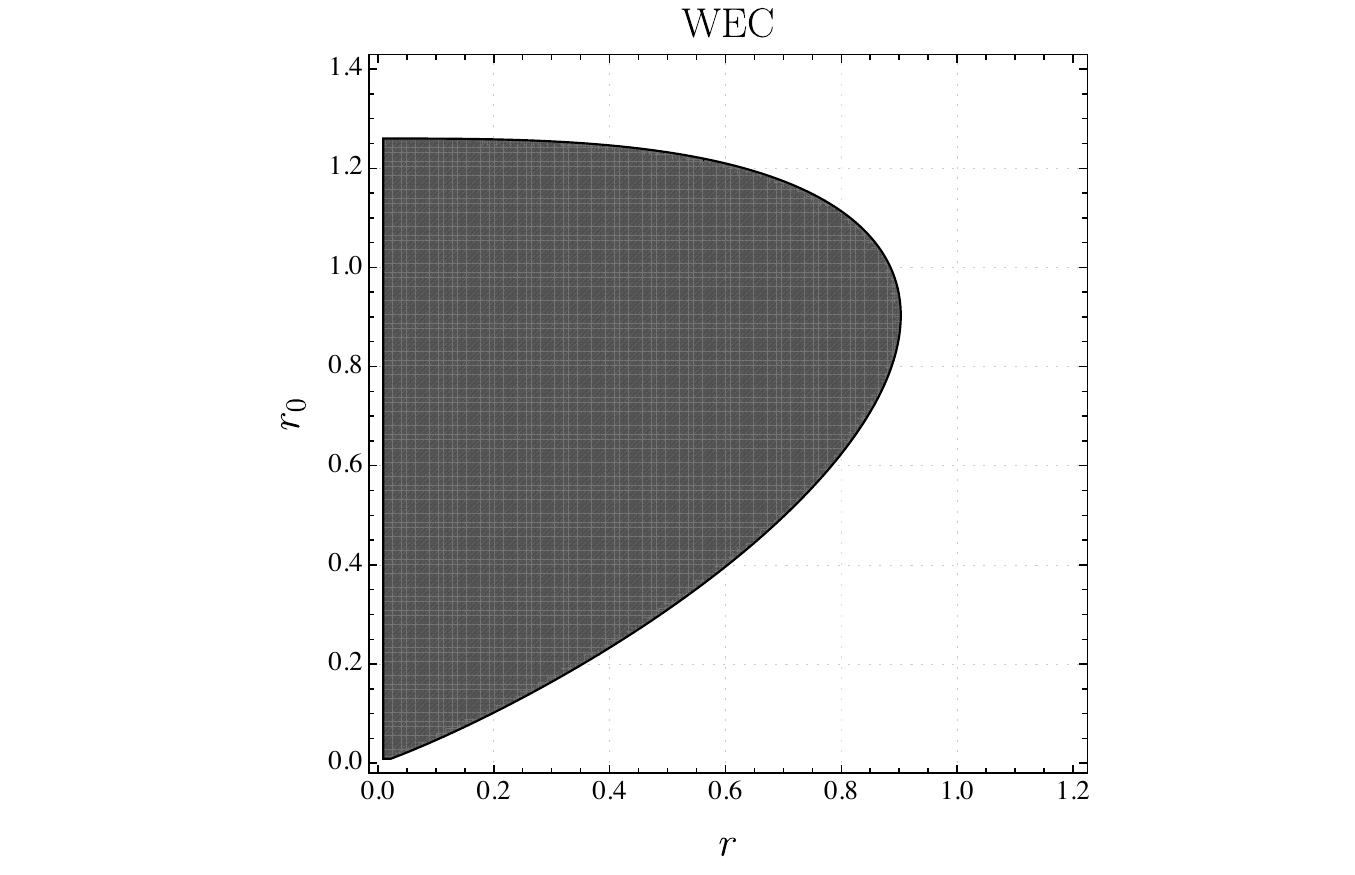}
\caption{Region in the $(r,r_0)$ parameter space where the weak energy condition is satisfied for $M=1$ and $\ell=1$, considering $0.01\leq r\leq1.2$ and $0.001\leq r_0\leq1.4$.}
	\label{fig:wec}
\end{figure}

The region in parameter space where the WEC is satisfied is shown in Fig.~\ref{fig:wec}.

\subsection{Strong energy condition}

For the present anisotropic source, the combinations
$\rho_{\mathrm{eff}}+p_{r,\mathrm{eff}}$ and
$\rho_{\mathrm{eff}}+p_{\perp,\mathrm{eff}}$ have already been shown to be nonnegative. The remaining nontrivial requirement of the strong energy condition (SEC) is therefore
\begin{equation}\label{eq:sec-definition}
	\rho_{\mathrm{eff}}
	+p_{r,\mathrm{eff}}
	+2p_{\perp,\mathrm{eff}}
	\geq0.
\end{equation}

Since $p_{r,\mathrm{eff}}=-\rho_{\mathrm{eff}}$, this condition is determined entirely by the transverse pressure. Substituting its explicit form gives
\begin{equation}\label{eq:sec}
	\frac{2M}{r_0^3}
	e^{-r^3/r_0^3}
	\left(
	\frac{3r^3}{2r_0^3}-1
	\right)
	+\frac{1}{\ell^2}
	\geq0.
\end{equation}

Near the origin, the sign of Eq.~\eqref{eq:sec} depends on the relative contribution of the Dymnikova matter distribution and the cosmological term. For $2M\ell^2>r_0^3$, the SEC may be violated in the central region, while for increasing $r$ the condition can be restored as the matter-dependent contribution changes sign and becomes exponentially suppressed.

\begin{figure}[htbp]
	\centering
	\includegraphics[width=\columnwidth]{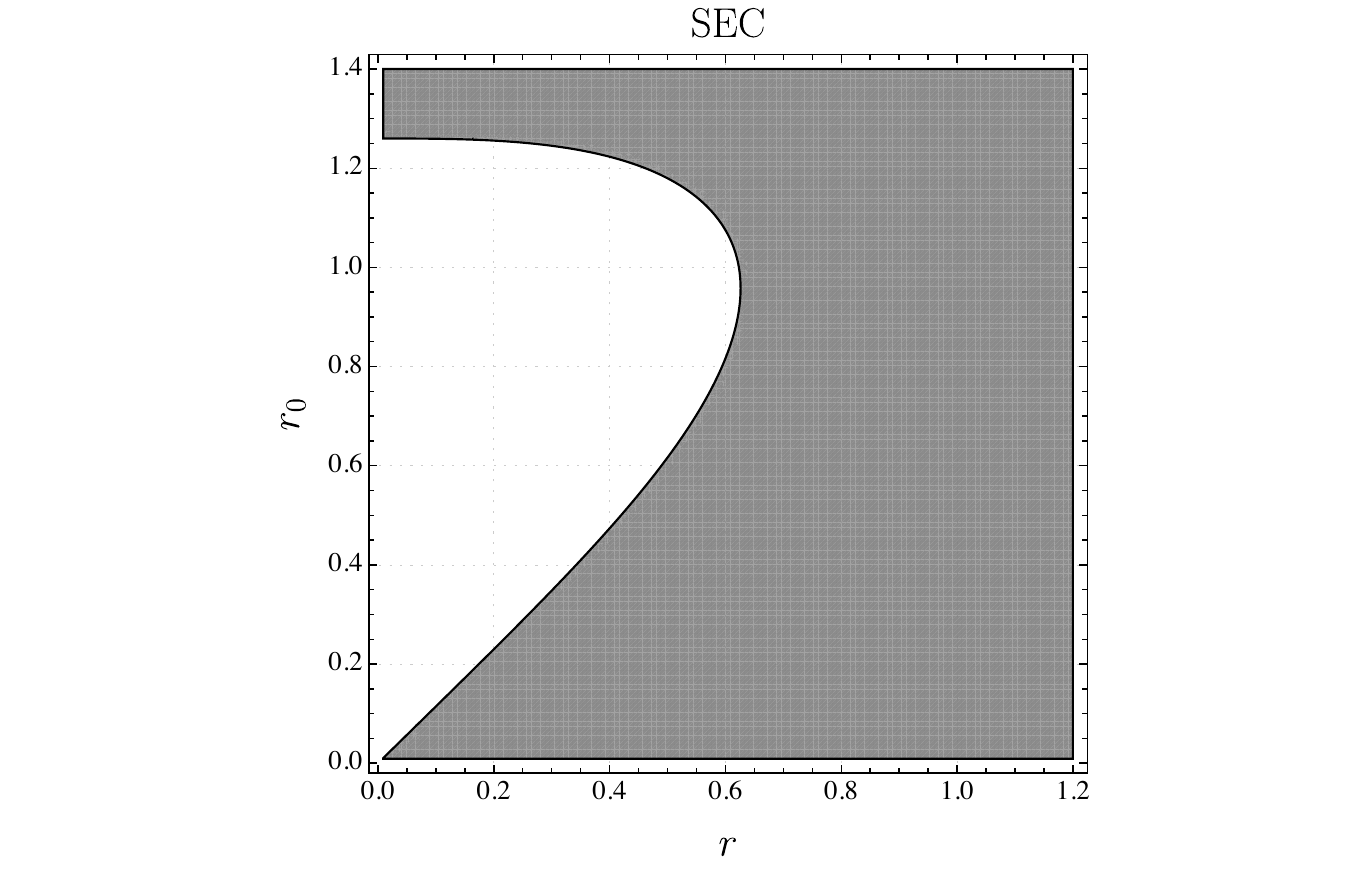}
\caption{Region in the $(r,r_0)$ parameter space where the strong energy condition is satisfied for $M=1$ and $\ell=1$, considering $0.01\leq r\leq1.2$ and $0.001\leq r_0\leq1.4$.}
	\label{fig:sec}
\end{figure}

At large radial distances, the exponential term vanishes and the cosmological contribution dominates Eq.~\eqref{eq:sec}. The corresponding region in parameter space is shown in Fig.~\ref{fig:sec}.

\begin{figure}[htbp]
	\centering
	\includegraphics[width=\columnwidth]{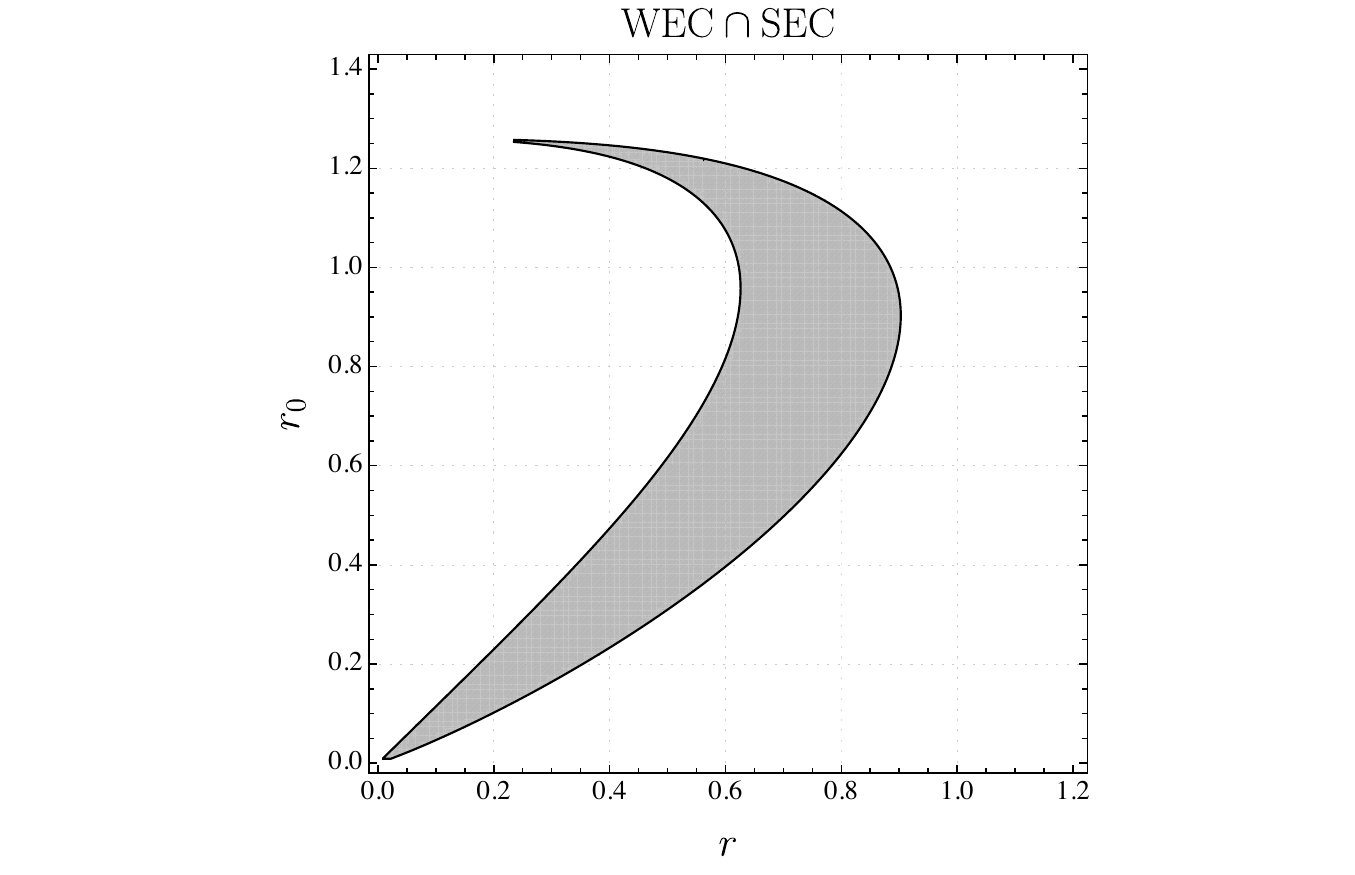}
	\caption{Region in the $(r,r_0)$ parameter space where the weak and strong energy conditions are simultaneously satisfied for $M=1$ and $\ell=1$, considering $0.01\leq r\leq1.2$ and $0.001\leq r_0\leq1.4$.}
	\label{fig:wec-sec}
\end{figure}

The WEC and SEC therefore impose different restrictions on the parameter space. The region where both conditions are simultaneously satisfied is shown in Fig.~\ref{fig:wec-sec}.

These regions characterize the effective energy density and pressures, whereas the event horizon is determined independently by the positive roots of the metric function $f(r)$.

\section{Event horizons and singularity structure}\label{sec:horizons-singularities}

Having characterized the metric function and its limiting behavior, we now examine the horizon and curvature structure of the solution. We first determine the event horizon from the positive zeros of $f(r)$ and investigate the possible existence of a finite-radius extremal configuration. The regularity of the spacetime will then be analyzed through the corresponding curvature invariants.

\subsection{Event horizon}

The event horizon is determined by the positive roots of the metric function, $	f(r_h)=0.$ For the present solution, this condition takes the form\footnote{The horizon equation in Eq.~(\ref{eq:horizon}) also allows an analytical characterization of the existence of positive roots. A positive event-horizon radius exists for $2M\ell^2>r_0^3$, which is equivalent to $\ell_{\mathrm{eff}}^{-2}<0$. For $2M\ell^2\leq r_0^3$, no positive event-horizon radius is obtained.}
\begin{equation}\label{eq:horizon}
	r_h^3
	+2M\ell^2 e^{-r_h^3/r_0^3}
	-2M\ell^2
	=0.
\end{equation}

Unlike the corresponding Lemos black string, the exponential term introduced by the Dymnikova matter distribution makes Eq.~\eqref{eq:horizon} transcendental. Consequently, the horizon radius cannot in general be written in a simple closed form and must be determined numerically once $M$, $\ell$, and $r_0$ are specified.

The parameter $r_0$ controls the characteristic scale of the matter distribution and enters directly into the horizon equation through the exponential term. Its variation therefore modifies the location of the positive root and, depending on the remaining parameters, the corresponding horizon structure.

For fixed $M$ and $\ell$, the dependence of the event-horizon radius on $r_0$ is summarized in Table~\ref{tab:horizon-radius}.

\begin{table}[htbp]
	\caption{Event-horizon radius for representative values of $r_0$, with $M=0.9$ and $\ell=1$.}
	\label{tab:horizon-radius}
	\centering
	\begin{tabular}{cc}
		\toprule
		$r_0$ & $r_h$\\
		\midrule
		0.2 & 1.2164\\
		0.4 & 1.2164\\
		0.6 & 1.2163\\
		0.8 & 1.2027\\
		1.0 & 1.0965\\
		\bottomrule
	\end{tabular}
\end{table}

We may also examine whether the Dymnikova regularization allows for a finite-radius extremal configuration. Extremality requires the simultaneous conditions
\begin{equation}
	f(r_e)=0,
	\qquad
	f'(r_e)=0.
\end{equation}
Combining these equations gives
\begin{equation}
	e^{r^{3}_{e}/r^{3}_{0}}
	=
	\frac{r^{3}_{e}}{r^{3}_{0}}+1.
\end{equation}
Since $e^x>1+x$ for every $x>0$, this relation has no solution with $r_e>0$. Equality occurs only at $r_e=0$. Therefore, the Dymnikova regularization does not introduce a finite-radius extremal horizon.

\subsection{Singularity structure}

We now examine the curvature invariants of the spacetime in order to characterize the behavior of the geometry at the origin. In particular, the Ricci and Kretschmann scalars provide a direct way of assessing whether the Dymnikova matter distribution is sufficient to soften the curvature behavior usually associated with the central region.

For the metric in Eq.~\eqref{eq:metric}, the Ricci scalar can be written directly in terms of the metric function as
\begin{equation}
	R=-f''-\frac{4f'}{r}-\frac{2f}{r^2}.
\end{equation}
Substituting the solution obtained previously gives
\begin{equation}
	R=-\frac{12}{\ell^2}+\frac{6M}{r_0^3}
	\left(4-\frac{3r^3}{r_0^3}\right)e^{-r^3/r_0^3}.
\end{equation}
The first term corresponds to the AdS contribution, while the second term encodes the modification generated by the Dymnikova matter distribution. The exponential factor suppresses this contribution at large radial distances, whereas near the origin it remains finite.

Using the effective curvature scale introduced previously, the limiting value of the Ricci scalar at the origin can be written as
\begin{equation}
	\lim_{r\to0}R=-\frac{12}{\ell_{\mathrm{eff}}^2}.
\end{equation}
Hence, the Ricci scalar remains finite at $r=0$. This indicates that the central behavior differs from that of the corresponding singular black-string geometry.

A more stringent characterization of the curvature is provided by the Kretschmann scalar, which contains the full quadratic contraction of the Riemann tensor. For the metric considered here, it takes the general form
\begin{equation}\label{eq:K}
	K=R_{\mu\nu\alpha\beta}R^{\mu\nu\alpha\beta}
	=(f'')^2+4\left(\frac{f'}{r}\right)^2
	+4\left(\frac{f}{r^2}\right)^2.
\end{equation}

Substituting the metric function into Eq.~\eqref{eq:K}, the resulting expression can be written as
\begin{equation}
	K=\frac{24}{\ell^{4}}+\frac{12Me^{-r^{3}/r^{3}_{0}}\left(6r^{3}-8r^{3}_{0}\right)}{\ell^{2}r^{6}_{0}}+\frac{12M^{2}e^{-2r^{3}/r^{3}_{0}}}{r^{6}r^{12}_{0}}\mathcal{K},
\end{equation}
where the radial dependence of the last contribution has been collected into the auxiliary quantity
\begin{equation}
	\begin{aligned}
	\mathcal{K}&=27r^{12}+4r^{12}_{0}\left(e^{r^{3}/r^{3}_{0}}-1\right)^{2}\\&-8r^{3}r^{9}_{0}\left(e^{r^{3}/r^{3}_{0}}-1\right)-12r^{6}r^{6}_{0}\left(e^{r^{3}/r^{3}_{0}}-2\right).
	\end{aligned}
\end{equation}

Although inverse powers of $r$ appear explicitly in this expression, these terms do not lead to a divergence at the center. After taking the limit, the potentially singular contributions cancel and the Kretschmann scalar approaches the finite value
\begin{equation}
	\lim_{r\to0}K=\frac{24}{\ell_{\mathrm{eff}}^4}.
\end{equation}

The finiteness of both curvature invariants indicates that no divergence appears in the Ricci and Kretschmann scalars at the origin. At the same time, the exponential corrections vanish asymptotically, so the solution approaches its AdS black-string behavior at large radial distances.

\section{Thermodynamics}\label{sec:thermodynamics}

We now investigate the thermodynamic properties associated with the event horizon of the Dymnikova black string. The horizon condition allows the mass parameter to be expressed in terms of the horizon radius, providing a convenient parametrization of the thermodynamic quantities. We first analyze the Hawking temperature and its limiting behavior, and then study the heat capacity as a measure of the local thermal response of the solution.

\subsection{Temperature}

For a static geometry of the form considered here, the Hawking temperature is determined by the surface gravity at the event horizon. In units $\hbar=k_B=1$, it is given by
\begin{equation}
	T\left(r_{h}\right)=\frac{f^{'}\left(r_{h}\right)}{4\pi}.
\end{equation}
Evaluating the derivative of the metric function at $r=r_h$ gives
\begin{equation}\label{eq:T}
	T\left(r_{h}\right)=\frac{1}{4\pi}\left\{ \frac{2r_{h}}{\ell^{2}}+\frac{2M}{r^{2}_{h}}\left[1-e^{-r^{3}_{h}/r^{3}_{0}}\left(1+\frac{3r^{3}_{h}}{r^{3}_{0}}\right)\right]\right\}.
\end{equation}

The horizon condition $f(r_h)=0$ relates the mass parameter to the horizon radius according to
\begin{equation}\label{eq:mass-thermal}
	M(r_h)=\frac{r_h^3}{2\ell^2(1-e^{-r_h^3/r_0^3})}.
\end{equation}
Substituting Eq.~\eqref{eq:mass-thermal} into Eq.~\eqref{eq:T}, the temperature can be expressed entirely in terms of $r_h$, $r_0$, and $\ell$ as
\begin{equation}
	T\left(r_{h}\right)=\frac{3r_{h}}{4\pi\ell^{2}}\left[1-\frac{r^{3}_{h}}{r^{3}_{0}\left(e^{r^{3}_{h}/r^{3}_{0}}-1\right)}\right].
\end{equation}

The characteristic scale $r_0$ controls the departure from the usual large-horizon behavior. The two limiting regimes are summarized in Table~\ref{tab:temperature-asymptotics}. For $r_h\ll r_0$, the temperature is strongly suppressed and grows as $r_h^4$. In contrast, when $r_h\gg r_0$, the exponential corrections become negligible and the temperature becomes linear in the horizon radius.

\begin{table}[htbp]
	\caption{Asymptotic behavior of the Hawking temperature in the regimes determined by the Dymnikova scale $r_0$.}
	\label{tab:temperature-asymptotics}
	\centering
	\begin{tabular}{ll}
		\toprule
		Regime & $T(r_h)$\\
		\midrule
		$r_{h}\ll r_{0}$ &
		$\displaystyle T\left(r_{h}\right)\simeq
		\frac{3r^{4}_{h}}{8\pi r^{3}_{0}\ell^{2}}$\\[6pt]
		$r_{h}\gg r_{0}$ &
		$\displaystyle T\left(r_{h}\right)\simeq
		\frac{3r_{h}}{4\pi\ell^{2}}$\\
		\bottomrule
	\end{tabular}
\end{table}

\begin{figure}[htbp]
	\centering
	\includegraphics[width=0.92\columnwidth]{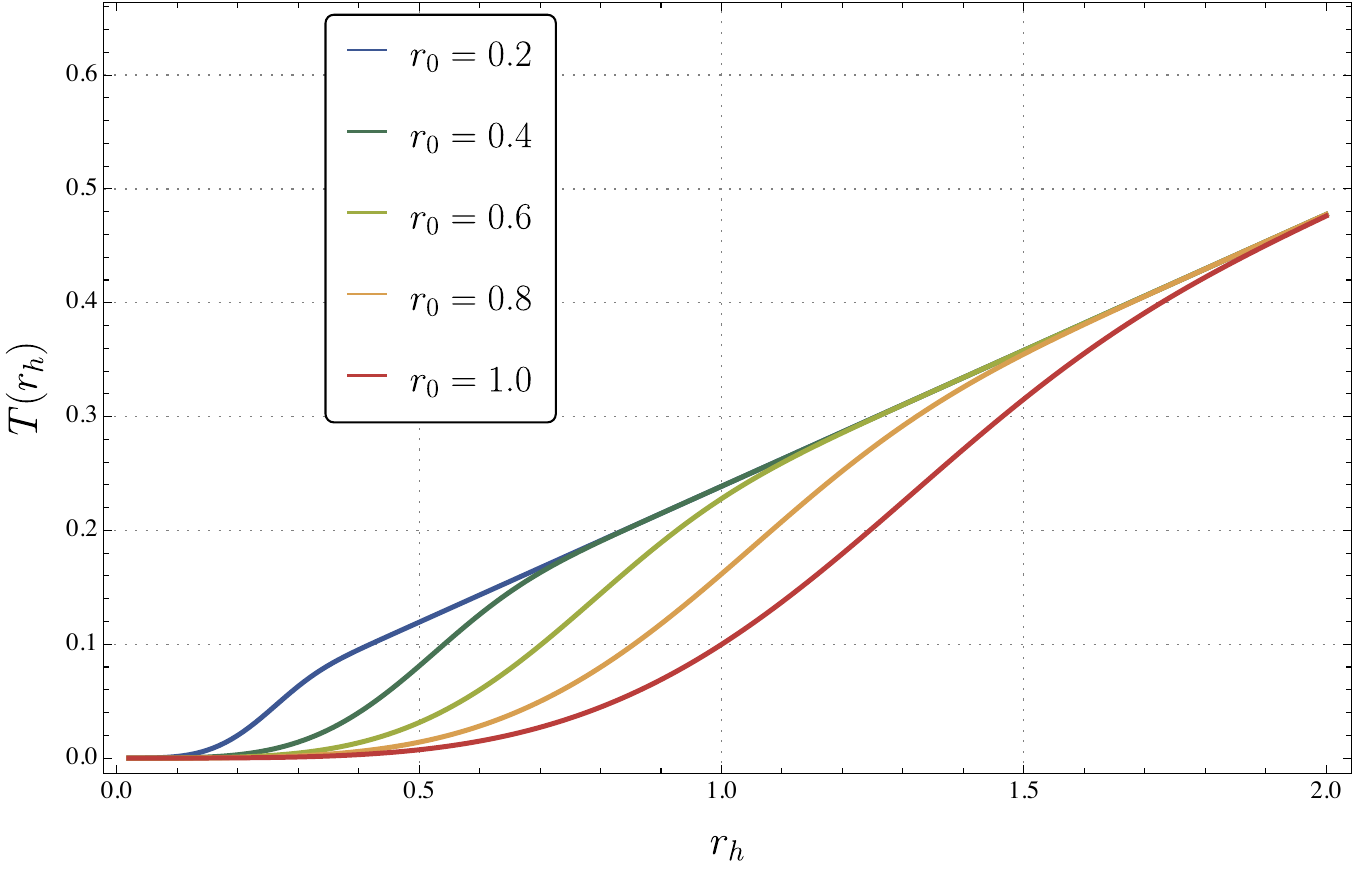}
	\caption{Hawking temperature $T(r_h)$ as a function of the horizon radius for $M=0.9$ and $\ell=1$, considering the representative values $r_0=0.2$, $0.4$, $0.6$, $0.8$, and $1.0$.}
	\label{fig:temperature}
\end{figure}

The complete dependence of the Hawking temperature on the horizon radius, including the transition between these two regimes, will be illustrated in Fig.~\ref{fig:temperature}.

\subsection{Heat capacity}

The local thermal response of the solution can be characterized by the heat capacity. Keeping $r_0$ and $\ell$ fixed, we define
\begin{equation}\label{eq:C-definition}
	C=\left(\frac{\partial M}{\partial T}\right)_{r_0,\ell}.
\end{equation}
Since both the mass and the temperature are parametrized by the horizon radius, the heat capacity can be evaluated from the ratio
of their derivatives with respect to $r_h$. Differentiating Eq.~\eqref{eq:mass-thermal} gives
\begin{equation}\label{eq:dM}
	\frac{dM}{dr_h}=\frac{3r_h^2}{2\ell^2}
	\frac{1-(1+r_h^3/r_0^3)e^{-r_h^3/r_0^3}}
	{(1-e^{-r_h^3/r_0^3})^2}.
\end{equation}

The corresponding derivative of the temperature is
\begin{equation}\label{eq:dT}
	\begin{aligned}
		\frac{dT}{dr_{h}}=\frac{3}{4\pi\ell^{2}}\left[
		1-\frac{1}{r^{3}_{0}}\frac{4r^{3}_{h}}
		{\left(e^{r^{3}_{h}/r^{3}_{0}}-1\right)}
		+\frac{3r^{6}_{h}}{r^{6}_{0}}
		\frac{e^{r^{3}_{h}/r^{3}_{0}}}
		{\left(e^{r^{3}_{h}/r^{3}_{0}}-1\right)^{2}}
		\right].
	\end{aligned}
\end{equation}

Taking the ratio of Eqs.~\eqref{eq:dM} and \eqref{eq:dT} yields
\begin{equation}\label{eq:C}
	C\left(r_{h}\right)=\frac{2\pi r^{2}_{h}\left(1-\frac{r^{3}_{h}/r^{3}_{0}}{\left(e^{r^{3}_{h}/r^{3}_{0}}-1\right)}\right)}{\left(1-e^{-r^{3}_{h}/r^{3}_{0}}\right)\left(1-\frac{4r^{3}_{h}/r^{3}_{0}}{\left(e^{r^{3}_{h}/r^{3}_{0}}-1\right)}+\frac{3r^{6}_{h}/r^{6}_{0}e^{r^{3}_{h}/r^{3}_{0}}}{\left(e^{r^{3}_{h}/r^{3}_{0}}-1\right)^{2}}\right)}.
\end{equation}

For $r_h>0$, the denominator of Eq.~\eqref{eq:C} does not vanish, and therefore the heat capacity has no finite-radius divergences. Its limiting behavior is summarized in Table~\ref{tab:heat-capacity-asymptotics}.

\begin{table}[htbp]
	\caption{Asymptotic behavior of the heat capacity in the regimes determined by the Dymnikova scale $r_0$.}
	\label{tab:heat-capacity-asymptotics}
	\centering
	\begin{tabular}{ll}
		\toprule
		Regime & $C(r_h)$\\
		\midrule
		$r_{h}\ll r_{0}$ &
		$\displaystyle C\left(r_{h}\right)\simeq
		\frac{\pi r^{3}_{0}}{2r_{h}}$\\[6pt]
		$r_{h}\gg r_{0}$ &
		$\displaystyle C\left(r_{h}\right)\simeq
		2\pi r^{2}_{h}$\\
		\bottomrule
	\end{tabular}
\end{table}

In the small-horizon regime, the heat capacity is positive and behaves as $C\propto r_h^{-1}$, diverging in the limiting case $r_h\to0^{+}$. This behavior accompanies the strong suppression of the Hawking temperature in the same regime. For $r_h\gg r_0$, the heat capacity approaches the quadratic behavior $C\simeq2\pi r_h^2$, recovering the large-horizon scaling of the asymptotic black string geometry. The growth of $C$ in this regime is asymptotic and is not associated with a finite-radius pole.

The exact expression in Eq.~\eqref{eq:C} provides a smooth interpolation between these two limiting regimes. In particular, no additional divergence occurs for finite $r_h>0$. 

\section{Conclusions}\label{Conclusions}

In this work, we constructed a static black string solution of Einstein's equations with a negative cosmological constant sourced by a Dymnikova-type anisotropic matter distribution. The resulting geometry approaches the static Lemos black string at large radial distances, while the matter profile modifies the behavior of the metric in the central region.

Near the origin, the leading behavior of the geometry is governed by the effective curvature parameter
$\ell_{\mathrm{eff}}^{-2}=\ell^{-2}-2M/r_0^3$.
Depending on the relative magnitude of the cosmological and matter contributions, the leading quadratic behavior of the metric can be AdS-like, vanish, or become de Sitter-like. The Ricci and Kretschmann scalars remain finite at the origin, with limiting values determined by the same effective curvature parameter. Thus, at the level of these curvature invariants, no divergence appears in the central region.

We also examined the weak and strong energy conditions by including the cosmological constant as an effective contribution to the energy density and pressures. The WEC is satisfied within a finite radial region when $2M\ell^2>r_0^3$, whereas for $2M\ell^2\leq r_0^3$ no positive radial interval satisfying the WEC is obtained. The SEC displays a different radial behavior and, depending on the parameters, can be violated near the central region and subsequently restored as the matter-dependent contribution changes sign and becomes exponentially suppressed. The corresponding regions in parameter space, including the region where both conditions are simultaneously satisfied, were examined numerically.

The event horizon is determined by a transcendental equation generated by the exponential Dymnikova contribution. Although its location must in general be determined numerically for fixed values of $M$, $\ell$, and $r_0$, the existence of a positive horizon can also be characterized in terms of the model parameters. Within the present analysis, horizon-bearing configurations correspond to the branch with $\ell_{\mathrm{eff}}^{-2}<0$, associated with the de Sitter-like central behavior. The extremality condition admits no solution at finite positive radius, so no finite-radius extremal configuration arises.

The thermodynamic analysis shows that the Dymnikova scale $r_0$ mainly affects the regime in which the horizon radius is comparable to or smaller than the characteristic matter scale. The Hawking temperature is strongly suppressed for $r_h\ll r_0$, whereas for $r_h\gg r_0$ it approaches the linear behavior associated with the asymptotic black-string geometry. The heat capacity similarly interpolates between a small-horizon regime, where it behaves as $r_h^{-1}$, and the large-horizon behavior $C\simeq2\pi r_h^2$, with no additional divergence occurring at finite positive $r_h$.

Overall, the construction provides a simple setting in which a Dymnikova-type matter profile can be incorporated into a static cylindrically symmetric geometry while preserving the asymptotic Lemos behavior and modifying the central curvature structure. A possible extension of the present analysis is the inclusion of rotation, which would allow one to investigate how the Dymnikova-type matter distribution affects the horizon structure, curvature properties, and thermodynamic behavior of rotating black strings.

\section{Acknowledgements}\label{Acknowledgements}
L.G.B. would like to thank the Coordenação de Aperfeiçoamento de Pessoal de Nível Superior – Brasil (CAPES) – Finance Code 001 for the financial support. L.G.B. also wishes to express gratitude to CENTRA for its hospitality. L.C.N.S would like to thank Conselho Nacional de Desenvolvimento Científico e Tecnológico - Brazil (CNPq) for financial support under Research Project No. 443769/2024-9 and Research Fellowship No. 314815/2025-2. C. C. Barros Jr. also thanks CNPq for financial support.

\bibliographystyle{unsrturl}
\bibliography{sample}

\end{document}